\pdfoutput=1

\documentclass[11pt]{article}

\usepackage[preprint]{acl}

\usepackage{times}
\usepackage{latexsym}

\usepackage[T1]{fontenc}

\usepackage[utf8]{inputenc}

\usepackage{microtype}

\usepackage{inconsolata}

\usepackage[utf8]{inputenc} 
\usepackage[T1]{fontenc}    
\usepackage{booktabs}       
\usepackage{amsfonts}       
\usepackage{nicefrac}       
\usepackage{microtype}      

\newif\ifrevise
\newif\ifrevisenew
\newif\iffinal

\revisetrue
\revisenewtrue
\finaltrue

\revisefalse
\revisenewfalse
\finalfalse

\usepackage{xspace}
\usepackage{multirow}
\usepackage{graphicx}
\usepackage{algorithm}
\usepackage{amsmath, amsthm, amssymb}
\usepackage{algorithmicx}
\usepackage[noend]{algpseudocode}
\usepackage{bm}
\usepackage[most]{tcolorbox}

\usepackage{longtable}
\usepackage{subfigure}
\usepackage{wrapfig}
\usepackage{tikz}
\usepackage{mdframed}
\usepackage{tabularx} 

\usepackage{cleveref}

\usepackage{pifont} 
\usepackage{array}
\newtcolorbox{colorquote}[1][]{
    boxrule=0.5pt,
    left=1pt,
    right=1pt,
    top=1pt,
    bottom=1pt,
    colback=black!5,
    colframe=black!55,
    notitle,
    enhanced,
    breakable,
}
\definecolor{myred}{RGB}{224,0,0}
\definecolor{myblue}{RGB}{46,117,182}
\definecolor{mygreen}{RGB}{83,130,53}
\definecolor{myyellow}{RGB}{191,144,0}
\usepackage{mdframed}
\usepackage{colortbl}
\newmdenv[linewidth=1pt, linecolor=blue, backgroundcolor=gray!20, roundcorner=10pt]{myframe}

\def\Snospace~{\S{}}

\newcounter{finding}

\title{From Effectiveness to Efficiency: Uncovering Linguistic Bias in Large Language Model-based Code Generation}

\author{
\textbf{Weipeng Jiang}\textsuperscript{1} ,
\textbf{Xuanqi Gao}\textsuperscript{1} ,
\textbf{Juan Zhai}\textsuperscript{2},
\textbf{Shiqing Ma}\textsuperscript{2}, \\
\textbf{Xiaoyu Zhang}\textsuperscript{1} ,
\textbf{Ziyan Lei}\textsuperscript{1}, 
\textbf{Chao Shen}\textsuperscript{1} 
\\
\normalsize{
\textsuperscript{1}Xi'an Jiaotong University,
\textsuperscript{2}University of Massachusetts Amherst}
\\
\normalsize{\texttt{\{\{lenijwp, gxq2000, zxy0927, l13201738997\}@stu, chaoshen@mail\}.xjtu.edu.cn}}
\\
\normalsize{ \texttt{\{juanzhai, shiqingma\}@umass.edu}}
}

\begin{document}

\maketitle
\begin{abstract}

Large Language Models (LLMs) have demonstrated promising capabilities for code generation. While existing benchmarks evaluate the correctness and efficiency of LLM-generated code, the potential linguistic bias - where code quality varies based on the natural language used to describe programming tasks - remains underexplored. 
In this paper, we aim to investigate this linguistic bias through the lens of English and Chinese.
To facilitate our investigation, we present a unified evaluation framework comprising a curated dataset of 52 Python programming questions with parallel bilingual task descriptions, automated correctness verification, and efficiency quantification tools based on runtime complexity estimation.
Based on this framework, we conduct the first empirical study towards the linguistic bias in LLM-generated code on eight popular LCGMs, as well as GPT-3.5-Turbo and GPT-4. 
We observe that these LCGM-generated code show different correctness on an average of 12\% bilingual programming tasks, where 39\% also exhibits diverse efficiency. 
Our findings indicate that LLMs commonly exhibit linguistic bias for code generation.
\end{abstract}

\section{Introduction}\label{sec:intro}


The rapid advancement of Large Language Models (LLMs) has revolutionized software engineering, demonstrating remarkable capabilities in code generation tasks ranging from simple code snippets~\cite{yin2017syntactic, bubeck2023sparks} to complete software development workflows~\cite{pudari2023copilot, qian2023communicative}. The increasing adoption of LLM for code generation has sparked significant research interest in evaluating the quality of generated codes.
Early evaluation benchmarks focused primarily on assessing code correctness across various programming scenarios, e.g., from function-level to project-level programming~\cite{chen2021evaluating, austin2021program}. 
Notable examples include HumanEval~\cite{chen2021evaluating}, HBPP~\cite{austin2021program} for basic programming task or competition-level problems, and DS-1000~\cite{lai2023ds} for data science applications. These benchmarks have provided valuable insights into LLMs' capabilities in generating functionally correct code. 
More recently, benchmarks like Effibench~\cite{huang2024effibench} have begun investigating the efficiency of LLM-generated code, recognizing its crucial role in developing scalable and sustainable software systems to meet the growing demands of the digital world.

\begin{figure*}[t]
  \centering
  \subfigure[Different Correctness]{
    \includegraphics[width=0.4\linewidth]{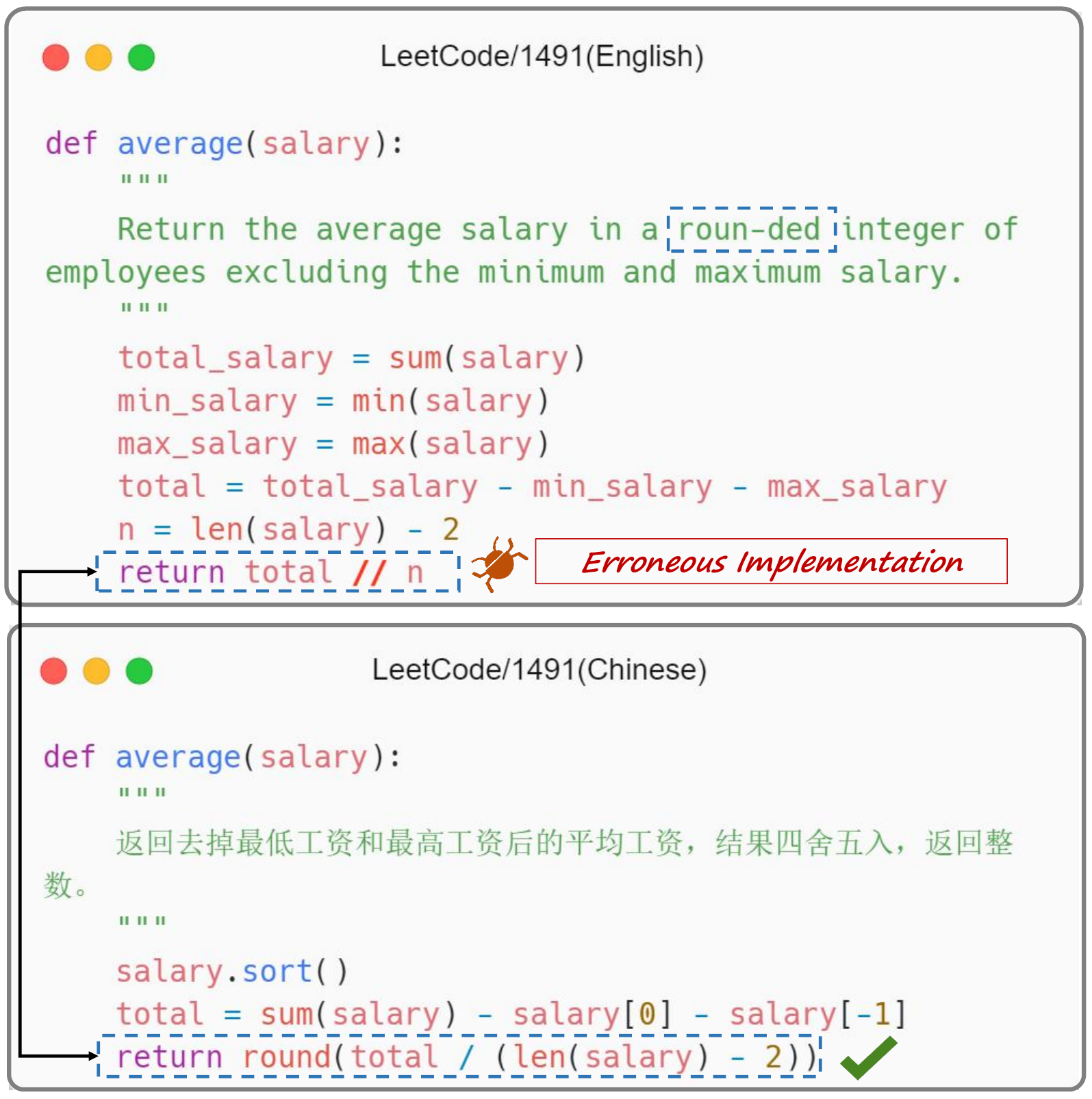}
  }
  \subfigure[Different Efficiency]{
    \includegraphics[width=0.4\linewidth]{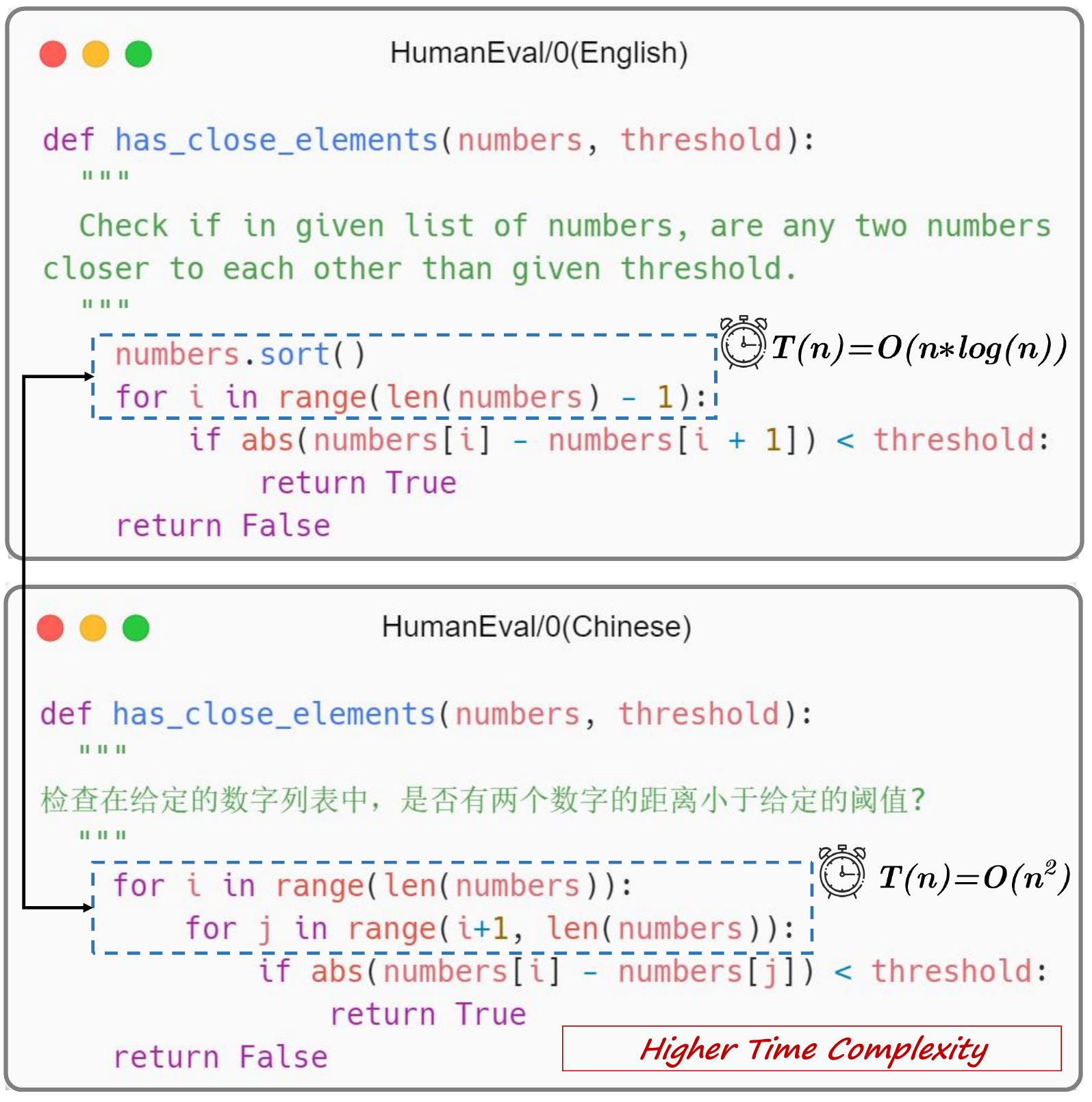}
  }
  \caption{Examples of linguistic bias in LLMs for code generation. The code generated by GPT-3.5 for the same programming task in English and Chinese exhibits differences in correctness and efficiency.}
  \label{fig:motivation}
\end{figure*}

However, a critical yet understudied aspect is the linguistic bias~\cite{guo2024bias} - the variation in code quality when programming tasks are described in different natural languages. This bias significantly impacts whether developers worldwide can equitably benefit from these technologies. 
We illustrate this concern with two examples in~\autoref{fig:motivation}. In case (a), when presented with a task to calculate the average salary, the LLM generates correct code for the Chinese description but fails for the English equivalent. Case (b) reveals a more subtle bias: while both English and Chinese inputs yield functionally correct code, the Chinese version employs nested loops with $O(n^2)$ complexity, whereas the English version utilizes the more efficient \(sort()\) function with $O(n \log n)$ complexity. Such efficiency disparities, though less apparent than correctness issues, can significantly impact the user experience in real-world applications and contribute to the concerning trend of degrading code quality on the Internet~\cite{CodingCopilot}.

While being intuitive, using existing code generation benchmarks directly for assessing linguistic bias in LLM-generated code from code correctness to efficiency has several limitations.
First, these benchmarks predominantly focus on programming questions exclusively in English, whereas evaluating linguistic bias inherently requires parallel corpora in at least two languages for meaningful comparison. 
Second, current benchmarks tend to evaluate either correctness or efficiency in isolation, lacking a unified framework for comparative analysis across both quality dimensions.
Finally, several benchmarks possess inherent shortcomings.
For example, HumanEval relies on a limited set of unit test cases to verify code correctness, which may be insufficient for thorough evaluation~\cite{liu2024your}, while Effibench assesses efficiency solely by measuring execution time or memory usage for the given input, which is highly sensitive to hardware configurations and execution environments, potentially leading to inconsistent performance evaluations across different systems.

This paper presents a unified evaluation framework for investigating linguistic bias in LLM-generated code, using English and Chinese as representative languages. Our framework is built upon a curated dataset of 52 high-quality bilingual programming question pairs, each accompanied by an automated test input generator and corresponding canonical solutions. We enhance correctness verification through comprehensive unit testing with large-scale sampled inputs. Additionally, we employ a profiling mechanism that analyzes execution time across varying input scales to determine code efficiency through complexity curve fitting. Through this framework, we systematically evaluate both the correctness and efficiency of LLM-generated code across different input languages.

Our comprehensive empirical study across 10 LLMs reveals significant disparities in code correctness and efficiency between English and Chinese programming questions. This linguistic bias persists even in larger models with enhanced programming capabilities, and varies notably with different prompting strategies. These findings demonstrate that LLMs do not yet consistently generate high-quality code across different input languages, highlighting a critical area for future research.
The main contributions of this paper are:
\begin{itemize}
\item We present the first systematic investigation of linguistic bias in LLM code generation across different natural languages.
\item We evaluate eight popular open-source LLMs and two commercial closed-source LLMs using our curated dataset. Our results show that 12\% of programming tasks exhibit different correctness across languages, while 39\% of correctly solved problems demonstrate efficiency disparities.
\item We provide our dataset and evaluation framework\footnote{\url{https://github.com/lenijwp/BilingualCodeEval}} to support future research in addressing linguistic bias in code generation.
\end{itemize}

\section{Related Work}
\label{sec:rw}

\textbf{LLM for Code Generation.} 
Large Language Models (LLMs) for code generation can be broadly categorized into two types. The first category comprises general-purpose LLMs, where state-of-the-art models also demonstrate remarkable code generation capabilities \cite{gpt3.5-oepnai, claude35,deepseekai2024deepseekv3technicalreport}.
The second category consists of code-specialized models, ranging from early dedicated architectures like CodeT5+~\cite{wang2021codet5}  to recent models fine-tuned from general-purpose LLMs, such as CodeFusion~\cite{singh2023codefusion}, CodeGen~\cite{Nijkamp2023codegen2}, StarCoder~\cite{li2023starcoder}, CodeGeeX~\cite{zheng2023codegeex}, CodeLlama\cite{roziere2023CodeLlama}, and DeepSeek-Coder\cite{guo2024deepseek}. 
We cover both categories in our evaluation.


\noindent
\textbf{Evaluation on LLM-generated Code.}
A set of benchmarks have been proposed to evaluate correctness and functionality of LLM-generate code, from function-level generation~\cite{chen2021evaluating, austin2021program} to cross-file code and project-level completion~\cite{ding2024crosscodeeval,yu2024codereval}.
Besides, researchers have also explored the robustness of LLMs~\cite{mastropaolo2023robustness,yang2023important,wang2022recode,li2023cctest,yang2024robustness}.
Beyond correctness, Effibench~\cite{huang2024effibench} developed a large-scale benchmark to evaluate the efficiency of code generated by LLMs.
Furthermore, Huang et al.~\cite{huang2023bias} and Ling et al.~\cite{ling2024evaluating} have investigated the social bias in LLM-generated code.
In this work, we delve into the linguistic bias present in LLM-driven code generation. 
\section{Evaluation Design}
\label{sec:design}


To systematically evaluate quality differences in LLM-generated code for inputs in various languages, 
we select English and Chinese for parallel analysis due to their global prevalence.
Furthermore, if considerable linguistic bias is present between these two widely used languages, it implies that this issue is likely amplified in low-resource languages. 
Our evaluation framework consists of the following steps:
\ding{182} We carefully select programming questions from existing benchmarks and conduct rigorous bilingual task description annotation. For each question, we collect canonical solutions and develop customized test input generators to automatically generate test cases of varying sizes for assessing both correctness and efficiency.
\ding{183} We collect generated code from multiple popular LLMs.
\ding{184} Using numerous test inputs from the input generators, we verify the correctness of the generated code by executing it against canonical solutions.
\ding{185} By measuring the average execution time of LLM-generated code across different input sizes, we estimate their time complexity and perform quantitative scoring.
\ding{186} We conduct fine-grained analysis comparing the correctness and efficiency of code generated by different LLMs for each question, enabling comprehensive linguistic bias analysis across models.

\subsection{Dataset Collection}
\label{sec:design_dataset}

We draw programming questions from existing benchmarks including HumanEval~\cite{chen2021evaluating}, MBPP~\cite{austin2021program}, and LeetCode~\cite{leetcode}. 
To ensure the quality of our test suite, we filter questions based on three criteria: 
\ding{182} tasks must be of moderate difficulty to avoid overly challenging problems;
\ding{183} tasks have multiple potential solutions to increase the likelihood of observing efficiency differences\footnote{We search online for candidate problems and verify that each selected problem has at least two reference solutions with different time complexities. Two software engineering experts independently collected the data, and a third author with programming competition experience validated their intersection.};
\ding{184} tasks must involve inputs with variable dimensionality (e.g., arrays or lists) so that execution time correlates with input size. 
This process yields 52 diverse programming questions covering diverse tasks, e.g., dynamic programming, string processing, and array enumeration. 
For each question, we collect reference solutions for correctness verification and create high-quality bilingual annotations.
All bilingual question pairs are carefully reviewed to ensure consistency in task descriptions and adherence to language conventions.


A key component of our evaluation framework is the automated test input generator designed for each question. 
It serves two crucial roles: 
First, it generates diverse test cases for unit testing by incorporating randomness, enabling thorough correctness validation of LLM-generated code. 
Secondly and more importantly, it can accept a size parameter $n$ that determines the input dimensionality/scale of generated test inputs (e.g., for array sorting task, $n$ can specify the array length). 
This design enables us to profile the execution time of LLM-generated code across varying input sizes, facilitating empirical estimation of average time complexity for efficiency comparison.
Detailed examples of the collected programming questions and our input generators can be found in \autoref{sec:appendix_examples}.

\begin{table*}[t] 
    \centering
    \small
    \caption{Summary of Evaluation Metrics} 
    \label{tab:metricssummary} 
    \begin{tabular}{llp{0.7\textwidth}} 
    \toprule 
    \textbf{Category} & \textbf{Metric} & \textbf{Description} \\ 
    \midrule 
    \multirow{4}{*}{Correctness} 
    & $\mathbf{CR_{en}}$ & \textbf{Correctness Rate in English}: the proportion of English programming questions for which the LLM generates correct code. \\ 
    & $\mathbf{CR_{zh}}$ & \textbf{Correctness Rate in Chinese}: the proportion of Chinese programming questions for which the LLM generates correct code. \\ 
    & $\mathbf{CR_{bi}}$ & \textbf{Correctness Rate in Bilingual}: the proportion of questions where the LLM produces correct code in both languages. \\ 
    & $\mathbf{CDR}$ & \textbf{Correctness Difference Rate}: the proportion of questions where the LLM generates correct code in only one language. \\ 
    \midrule 
    \multirow{3}{*}{Efficiency} 
    & $\mathbf{PAR_{en}}$ & \textbf{Performance Advantage Rate in English}: among questions correctly solved in both languages, the proportion where the English solution has a lower estimated time complexity. \\ 
    & $\mathbf{PAR_{zh}}$ & \textbf{Performance Advantage Rate in Chinese}: among questions correctly solved in both languages, the proportion where the Chinese solution has a lower estimated time complexity. \\ 
    & $\mathbf{PDR}$ & \textbf{Performance Difference Rate}: among questions correctly solved in both languages, the proportion where the estimated time complexities differ significantly. \\ 
    \bottomrule 
    \end{tabular} 
\end{table*}

\subsection{Correctness Verification}
\label{sec:design_correctness}

As discussed earlier, only a few test cases in existing benchmarks are often insufficient to guarantee thorough functionality correctness judgment, such a concern is also highlighted in recent studies~\cite{liu2024your}.
To ensure the adequacy of unit testing, we automatically generate larger-size test cases for correctness verification.
As aforementioned, our collected dataset includes an input generator, which can generate random inputs of a specified size. 
With the canonical solution \(S\), we can automatically synthesize reference outputs, i.e., ground truth. 
Therefore, we can conduct more comprehensive unit testing.
We draw inspiration from HumanEval's approach, which involves executing the code in a sandbox environment and verifying correctness by comparing the output of the LLM-generated code with the synthesized ground truth.
Furthermore, we filter out the functionally correct code. 
Prioritizing correctness verification lays a solid foundation for subsequent performance estimation, as discussing the performance merits of code with inherent functional errors lacks practical significance. 
\subsection{Performance Estimation}
\label{sec:design_performance}

Although existing frameworks, such as Effibench, provide methodologies for directly monitoring runtime and memory consumption of LLM-generated code under specific inputs, the inherent susceptibility of raw performance metrics to hardware configurations and environmental factors limits their reproducibility and reusability across different computing environments. To address this limitation, we propose estimating performance complexity by dynamically sampling execution times across varying input sizes, enabling us to derive an average execution time curve that offers a more robust and environment-independent assessment of algorithmic efficiency.
We leverage the input generator to systematically profile the execution time of LLM-generated code across different input scales. We implement an adaptive mechanism to determine the optimal maximum input size for evaluation (detailed in \autoref{sec:appendix_time_series}). Through uniform segmentation of the input range, we generate a sequence of input-size and execution-time pairs, with execution times averaged over multiple runs to ensure statistical reliability. This progressive sampling strategy effectively captures the relationship between input scale and computational cost while mitigating the impact of environmental variations.

\label{sec:design_performance}

\begin{table}[t]
    \centering
    \small
    \caption{Common time complexity types.}
    \begin{tabular}{cc}
    \hline
    Type         &  Expression  \\ \hline
    Constant     &  $time = a \, \text{(sec)}$      \\
    Logarithmic  & $time = a \log(n) + b \, \text{(sec)}$     \\
    Linear       & $time = a n + b \, \text{(sec)}$      \\
    Linearithmic & $time = a n \log(n) + b \, \text{(sec)}$     \\
    Quadratic    & $time = a n^2 + b \, \text{(sec)}$       \\
    Cubic        & $time = a n^3 + b \, \text{(sec)}$       \\
    Exponential  &  $time = a^n + b \, \text{(sec)}$\\ \hline
    \end{tabular}
    \label{tab:bigO}
\end{table}

To characterize the observed performance patterns, we consider seven prevalent time complexity classes, as illustrated in \autoref{tab:bigO}. We employ least squares regression analysis~\cite{draper1998applied} on the collected time series data to identify the best-fitting complexity pattern and corresponding mathematical expression. Furthermore, we establish a quantitative scoring system, assigning scores from 1 to 7 to each complexity class according to their order in \autoref{tab:bigO}, with lower scores indicating higher algorithmic efficiency. This scoring mechanism enables objective comparison of code implementations based on their empirically observed complexity patterns.

\subsection{Measurement Metrics}\label{sec:metrics}

To systematically evaluate the discrepancies in code generation capabilities of LLMs when handling bilingual versions of identical programming problems, we comprise seven carefully designed metrics: four metrics quantify the differences in code correctness across languages, while three metrics measure the variations in code execution efficiency (see \autoref{tab:metricssummary}). For detailed definitions and analytical methodologies of these metrics, please refer to ~\autoref{sec:appendix_metrics}.

\subsection{Target LLMs}

In our study, we evaluated a diverse set of LLMs, including both open-source and commercial LLMs. 
Our model selection prioritizes three key criteria: multilingual capabilities (English/Chinese), architectural diversity to capture linguistic bias generalizability, and computational feasibility (i.e., the model has most 34B parameters).
We selected five families of open-source LLMs: CodeGen2.5-7B~\cite{Nijkamp2023codegen2}, StarCoder-13B~\cite{li2023starcoder}, CodeGeeX-6.7B~\cite{zheng2023codegeex}, CodeLlama (7B/13B/34B)\cite{roziere2023CodeLlama}, and DeepSeek-Coder (7B/33B)\cite{guo2024deepseek}(shortened to DeepSeek). Additionally, we included two commercial closed-source models, GPT-3.5-Turbo and GPT-4~\cite{gpt3.5-oepnai}, resulting in a total of ten evaluated models. 
More details on the selected models can be found in \autoref{sec:appendix_models}.



\section{Results and Analysis}
\label{sec:eval}



\subsection{RQ1: Study on Correctness.}
\label{sec:rq1}

\begin{table*}[tbp]
    \centering
    \footnotesize
    \setlength{\tabcolsep}{2pt}
    \caption{Results of Measurement on Correctness and Efficiency Difference.}
    \footnotesize
    \begin{tabular}{cccccccccccccccc}
    \hline
     &  & \multicolumn{7}{c}{\(t=0.2\)} & \multicolumn{7}{c}{\(t=0.8\)} \\ 
    \cmidrule(lr){3-9} \cmidrule(lr){10-16}
     & \multirow{-2}{*}{Model} & CR\_{en} & CR\_{zh} & CR\_{bi} & CDR & PAR\_{en} & PAR\_{zh} & PDR & CR\_{en} & CR\_{zh} & CR\_{bi} & CDR & PAR\_{en} & PAR\_{zh} & PDR \\ \hline
    & CodeGeeX(6.7B) & \cellcolor[HTML]{D9EAD3}0.60 & 0.54 & 0.54 & 0.06 & 0.17 & 0.17 & 0.33 & 0.62 & \cellcolor[HTML]{FCE5CD}0.63 & 0.58 & 0.14 & \cellcolor[HTML]{D9EAD3}0.34 & 0.20 & \cellcolor[HTML]{FFFFFF}0.54 \\
    & CodeGen2.5(7B) & 0.52 & 0.52 & 0.46 & 0.12 & 0.20 & \cellcolor[HTML]{FCE5CD}0.23 & 0.43 & 0.46 & \cellcolor[HTML]{FCE5CD}0.48 & 0.40 & 0.25 & 0.11 & \cellcolor[HTML]{FCE5CD}0.25 & \cellcolor[HTML]{FFFFFF}0.36\\
    & CodeLlama(7B) & 0.50 & \cellcolor[HTML]{FCE5CD}0.54 & 0.46 & 0.12 & 0.13 & \cellcolor[HTML]{FCE5CD}0.27 & 0.40 & \cellcolor[HTML]{D9EAD3}0.52 & 0.50 & 0.46 & 0.17 & \cellcolor[HTML]{D9EAD3}0.21 & 0.14 & \cellcolor[HTML]{FFFFFF}0.34 \\
    & DeepSeek(13B) & \cellcolor[HTML]{D9EAD3}0.63 & 0.58 & 0.54 & 0.13 & 0.20 & \cellcolor[HTML]{FCE5CD}0.23 & 0.43 & \cellcolor[HTML]{D9EAD3}0.71 & 0.65 & 0.63 & 0.13 & \cellcolor[HTML]{D9EAD3}0.24 & 0.11 & \cellcolor[HTML]{FFFFFF}0.34 \\
    & StarCoder(13B) & 0.48 & \cellcolor[HTML]{FCE5CD}0.52 & 0.44 & 0.12 & 0.17 & \cellcolor[HTML]{FCE5CD}0.24 & 0.41 & \cellcolor[HTML]{D9EAD3}0.58 & 0.56 & 0.52 & 0.16 & \cellcolor[HTML]{D9EAD3}0.26 & 0.13 & \cellcolor[HTML]{FFFFFF}0.39 \\
    & CodeLlama(13B) & 0.46 & \cellcolor[HTML]{FCE5CD}0.56 & 0.44 & 0.13 & 0.17 & \cellcolor[HTML]{FCE5CD}0.30 & 0.47  & 0.54 & \cellcolor[HTML]{FCE5CD}0.58 & 0.50 & 0.12 & 0.12 & \cellcolor[HTML]{FCE5CD}0.22 & \cellcolor[HTML]{FFFFFF}0.25\\
    & DeepSeek(33B) & 0.60 & \cellcolor[HTML]{FCE5CD}0.62 & 0.54 & 0.13 & 0.15 & \cellcolor[HTML]{FCE5CD}0.29 & 0.44  & \cellcolor[HTML]{D9EAD3}0.60 & 0.58 & 0.54 & 0.10 & \cellcolor[HTML]{D9EAD3}0.24 & 0.21 & 0.45 \\
    \multirow{-8}{*}{} & CodeLlama(34B) & 0.54 & \cellcolor[HTML]{FCE5CD}0.60 & 0.48 & 0.17 & 0.09 & \cellcolor[HTML]{FCE5CD}0.32 & 0.41 & \cellcolor[HTML]{D9EAD3}0.67 & 0.56 & 0.54 & 0.15 & \cellcolor[HTML]{D9EAD3}0.25 & 0.22 & 0.47 \\ \hline
    & GPT-3.5 & 0.58 & \cellcolor[HTML]{FCE5CD}0.64 & 0.58 & 0.08 & 0.06 & \cellcolor[HTML]{FCE5CD}0.29 & 0.35 & \cellcolor[HTML]{D9EAD3}0.70 & 0.66 & 0.67 & 0.05 & \cellcolor[HTML]{D9EAD3}0.22 & 0.19 & 0.41 \\
    \multirow{-2}{*}{} & GPT-4 & \cellcolor[HTML]{D9EAD3}0.65 & 0.63 & 0.61 & 0.06 & 0.14 & \cellcolor[HTML]{FCE5CD}0.17  & 0.31 & \cellcolor[HTML]{D9EAD3}0.71 & 0.65 & 0.65 & 0.06 & 0.14 & \cellcolor[HTML]{FCE5CD}0.19 & 0.32 \\ \hline
    \multicolumn{1}{l}{} & Average & 0.56 & \cellcolor[HTML]{FCE5CD}0.58 & 0.51 & 0.11 & 0.15 & \cellcolor[HTML]{FCE5CD}0.25 & 0.40 & \cellcolor[HTML]{D9EAD3}0.61 & 0.59 & 0.55 & 0.13 & \cellcolor[HTML]{D9EAD3}0.21 & 0.19 & 0.39 \\ \hline
    \end{tabular}
    \label{tab:merged}
\end{table*}


\noindent\textbf{Design.} 
In this section, we aim to investigate the performance disparities of LLMs when generating correct code for both Chinese and English programming tasks. Given the stochastic nature of LLM code generation, we conduct 20 independent generation trials for each LLM on every programming question. An LLM is considered capable of solving a question correctly if at least one of the generated code snippets passes the correctness verification. 
Table \ref{tab:merged} demonstrates the evaluation results by measurements discussed in~\autoref{sec:metrics}.

\noindent\textbf{Results.}
Table \ref{tab:merged} presents the results based on the correctness verification, under varied temperatures ($t=0.2$ and $t=0.8$). 
Each row corresponds to a specific LLM, and the columns display various metrics.
Firstly, we find that \textit{LLMs demonstrate linguistic bias, producing correctness-inconsistent code between English and Chinese inputs in over 10\% of cases}.
The column $CDR$ quantifies where a model successfully generates correct code in one language while failing in the other. 
Our analysis reveals that across both temperature settings, LLMs exhibit correctness inconsistencies in more than 12\% of cases. Notably, commercial models GPT-3.5-Turbo and GPT-4 demonstrate superior problem-solving capabilities while maintaining the lowest linguistic bias among all evaluated models.

Furthermore, we observe that \textit{sampling temperature influences the linguistic bias}: at lower temperatures ($t=0.2$), LLMs favor Chinese inputs, while at higher temperatures ($t=0.8$), they demonstrate superior performance in English code generation.
The $\text{CR}_{en}$ and $\text{CR}_{zh}$, represent the proportion of correctly solved programming questions for English and Chinese inputs, respectively.
Our results reveals a significant improvement in English code generation correctness when increasing the temperature from 0.2 to 0.8 across most models ($p\text{-value} = 0.008$). 
In contrast, Chinese input performance shows no statistically significant improvement ($p\text{-value} = 0.26$).
The comparative advantage between languages, indicated by green highlighting for English superiority and orange for Chinese, demonstrates a clear temperature-dependent shift. At $t=0.2$, only two models show superior performance in English, while at $t=0.8$, this number doubles to four models, indicating a temperature-induced bias toward English-language code generation.

\subsection{RQ2: Study on Efficiency.}
\label{sec:rq2}



\noindent\textbf{Design.} 
In this research question, we consider only those programming tasks that LLMs solve correctly in both English and Chinese. For each task, we collect all correct solutions in both languages and select the most efficient one—i.e., the solution with the lowest complexity score. This enables a direct comparison of code efficiency across languages and allows us to assess the impact of language on the optimization of generated code.

\begin{figure}
    \centering     
    \includegraphics[width=0.99\linewidth]{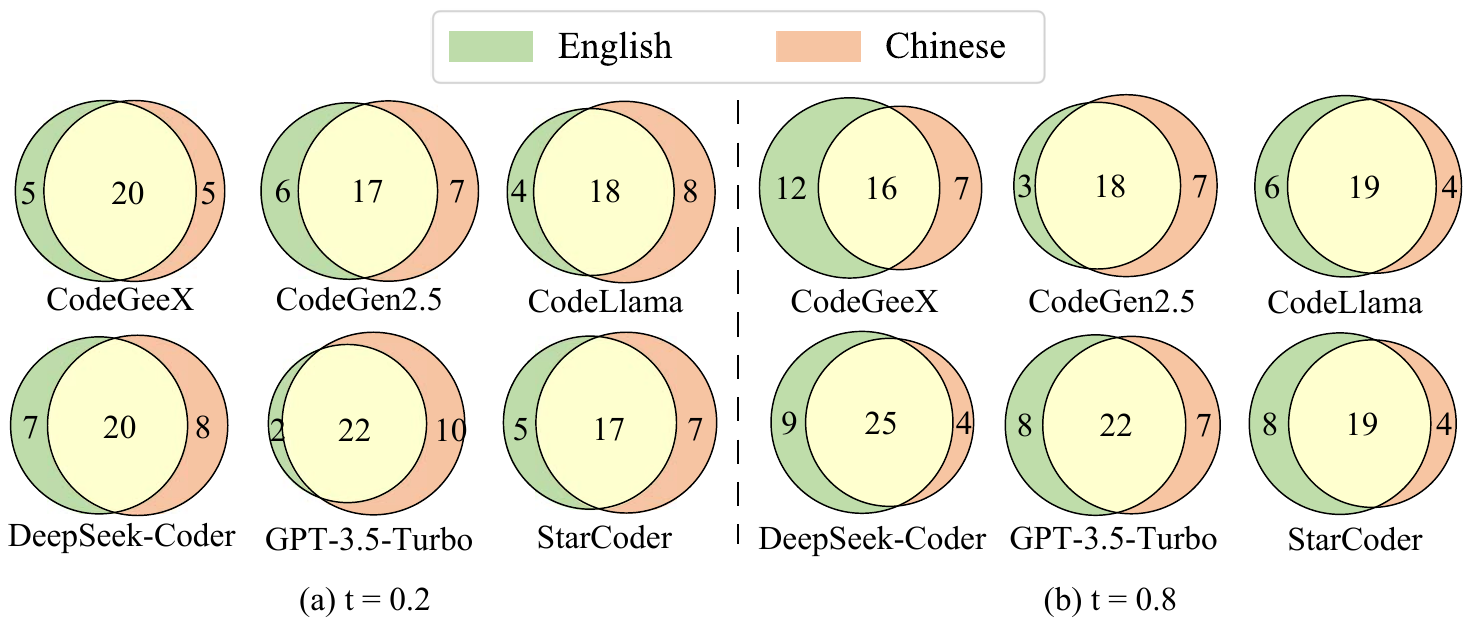}
    \caption{Venn Diagram of Efficiency Advantage.}\label{fig:perf_venn}
    \vspace{-10pt}
\end{figure}

\noindent\textbf{Results.}
We first identify that \textit{LLMs demonstrate a substantial linguistic bias in the efficiency of generated code}.
~\autoref{fig:perf_venn} illustrates the selected set of programming questions for each LLM, showing the number of questions where code efficiency is higher in one language compared to the other. 
The central overlap represents questions where code efficiency is similar for both languages. 
It is clear that, for each LLM, a substantial portion of codes shows better efficiency for one language over the other.
~\autoref{tab:merged} further presents the specific evaluation metrics.
$PDR$ provides a detailed measurement of this difference. 
Experimental results show that the average $PDR$ is over 39\% during various temperatures, which indicates that the efficiency of code generated by LLMs may have greater inconsistency than correctness, warranting attention.

\begin{figure}[!t]
    \centering
    \subfigure[t = 0.2]{
      \includegraphics[width=0.45\linewidth]{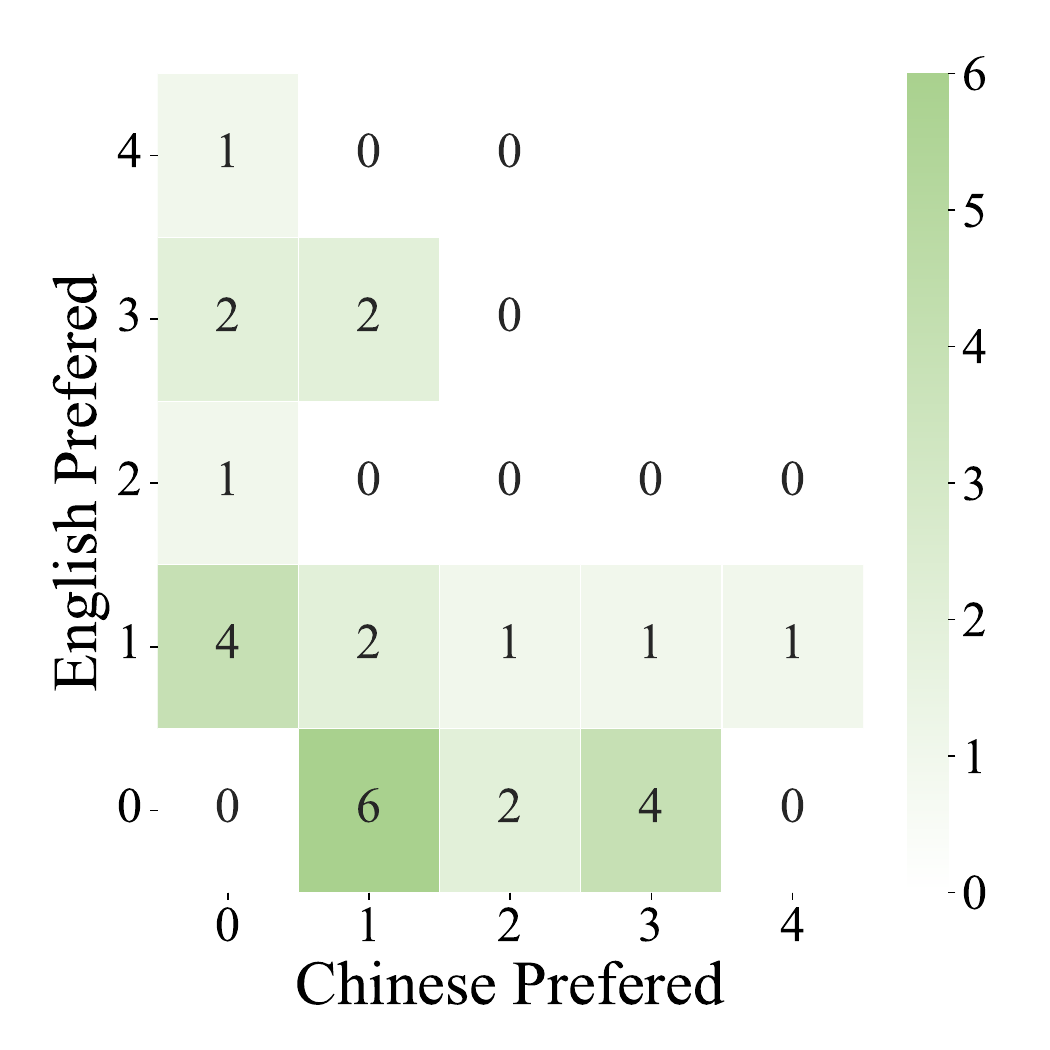}
    }
    \subfigure[t = 0.8]{
      \includegraphics[width=0.45\linewidth]{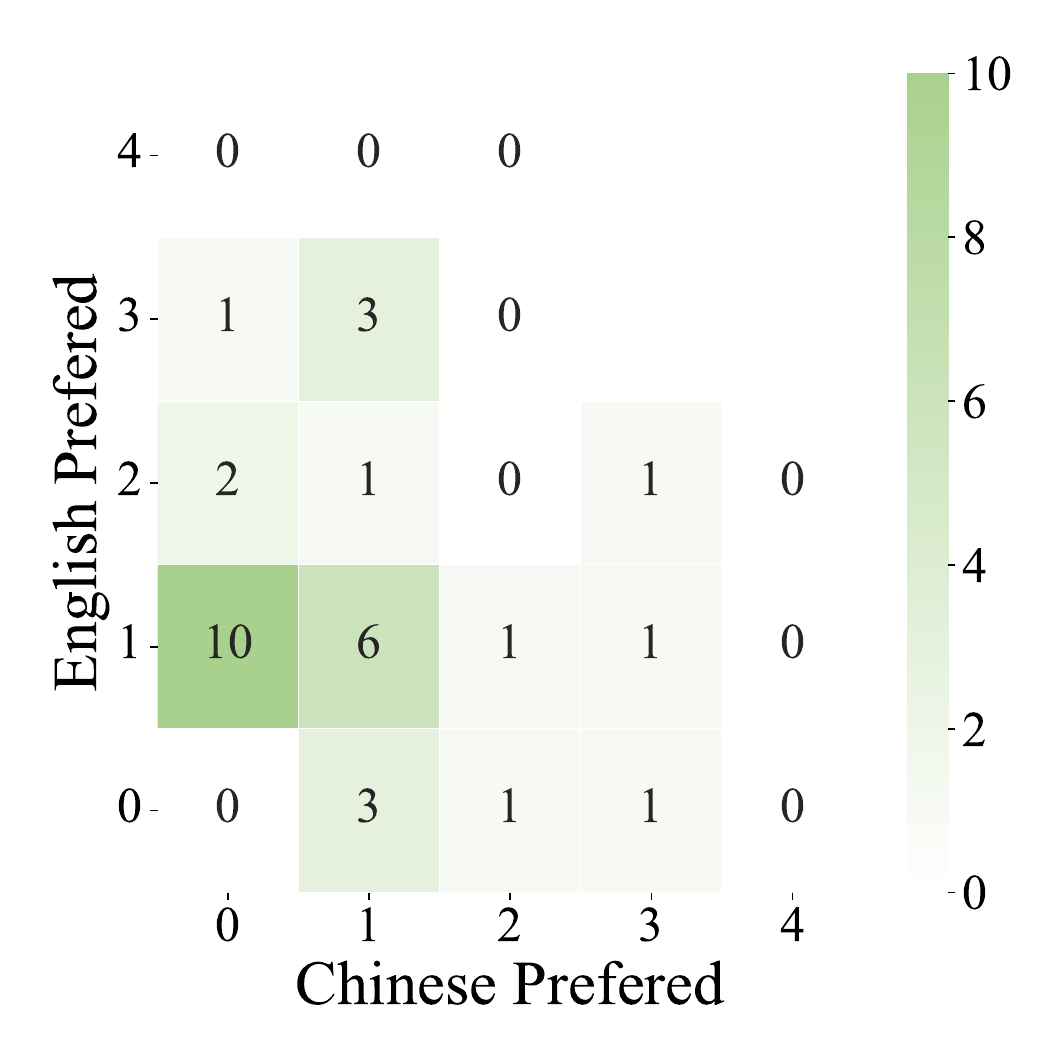}
    }
    \caption{The distribution of bilingual advantages of open-source LCGMs.}
    \label{fig:perf_heat}
\end{figure}

Meanwhile, we find that \textit{with rising temperatures, the advantage of generating more efficient code also shifts from Chinese to English inputs, which is similar with correctness}.
$PAR_{en}$ and $PAR_{zh}$ represent the proportion of programming questions where these LLMs gain an advantage in each language, as highlighted.
It can be observed that when $t=0.2$, five LLMs generate code with better efficiency for Chinese input; when $t=0.8$, five LLMs generate code with better efficiency for English inputs. In other words, as $t$ increases, the advantage shifts from Chinese to English, which is consistent with our observations in~\autoref{sec:rq1}.
~\autoref{fig:perf_heat} further illustrates the distribution of the efficiency advantages of the generated code for each LLM. The horizontal axis represents the number of LLMs that generate more efficient code in the Chinese version of the question, while the vertical axis represents the number of LLMs that generate more efficient code in the English version of the question. Experimental results show that when $t=0.2$, the darker shaded areas in the graph indicate a stronger preference for Chinese.



Additionally, we observe that \textit{larger LLMs may improve code generation capabilities but can also exacerbate linguistic bias, rather than mitigate it}.
At \( t=0.8 \), the linguistic bias is evident not only in the smallest model, CodeGeeX, but also in larger models such as CodeLlama-34B and GPT-3.5-Turbo. 
For CodeLlama, the $PDR$ initially decreases to its lowest value as the model size increases from 7B to 13B, but then rises to 0.47 as the size grows to 34B. This suggests that larger models, while potentially offering improved code generation capabilities, do not necessarily reduce bias and may even exhibit more pronounced bias.


\begin{figure*}[!t]
    \centering
    \includegraphics[width=\linewidth]{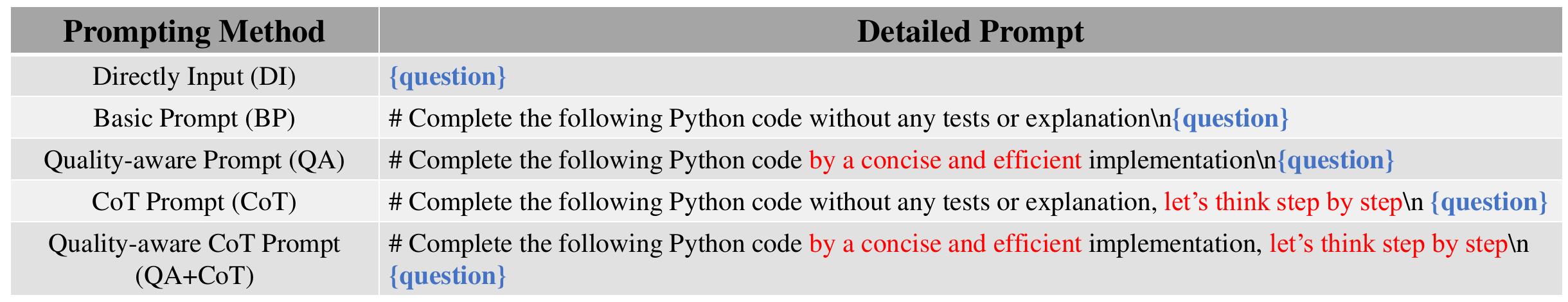}
    \caption{Different Prompting Methods.}

    \label{fig:prompting}
\end{figure*}


\begin{figure*}[!t]
    \centering
    \subfigure[Impact on Correctness]{
      \includegraphics[width=0.45\linewidth]{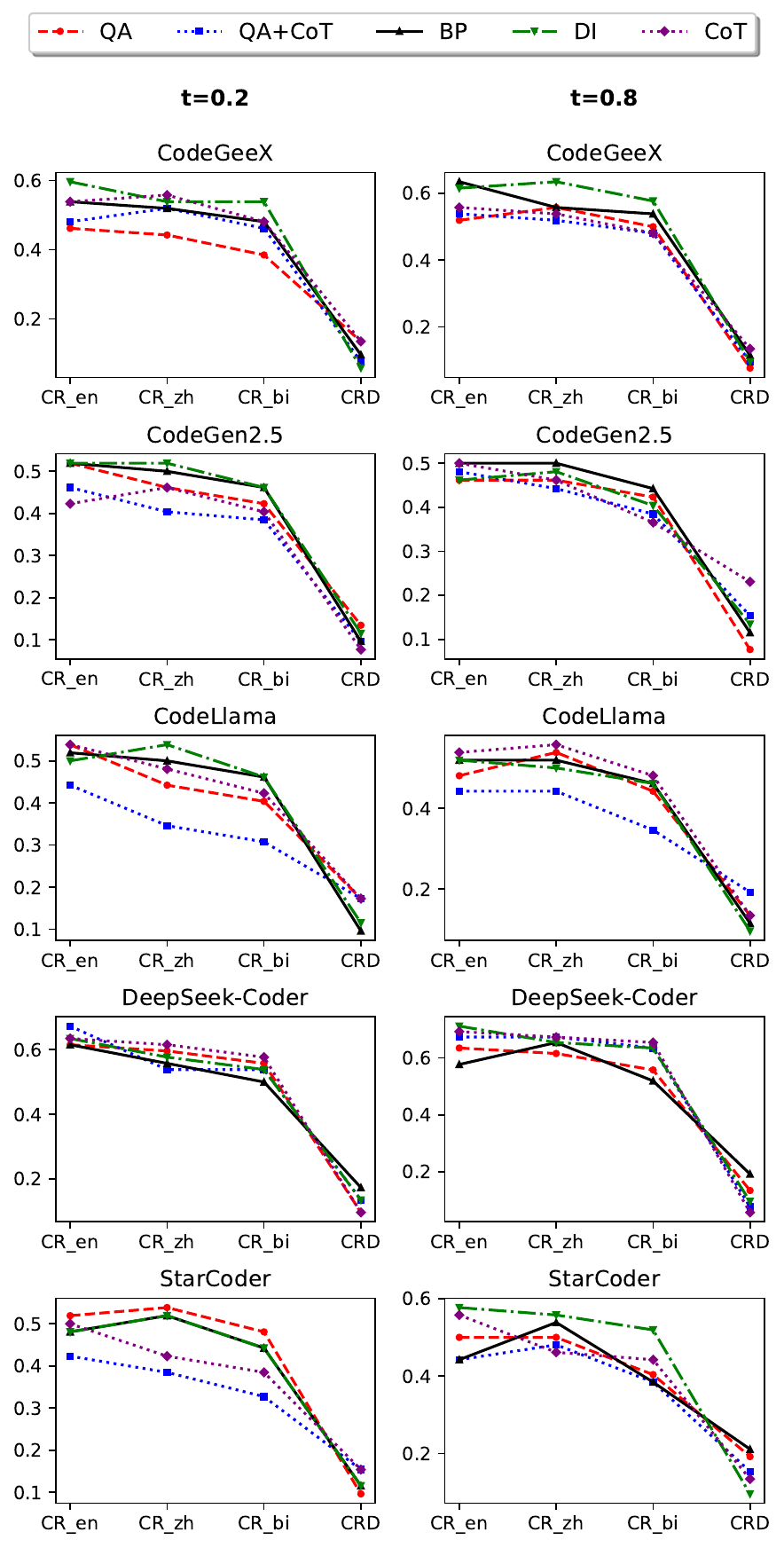}
    }
    \subfigure[Impact on Efficiency]{
      \includegraphics[width=0.45\linewidth]{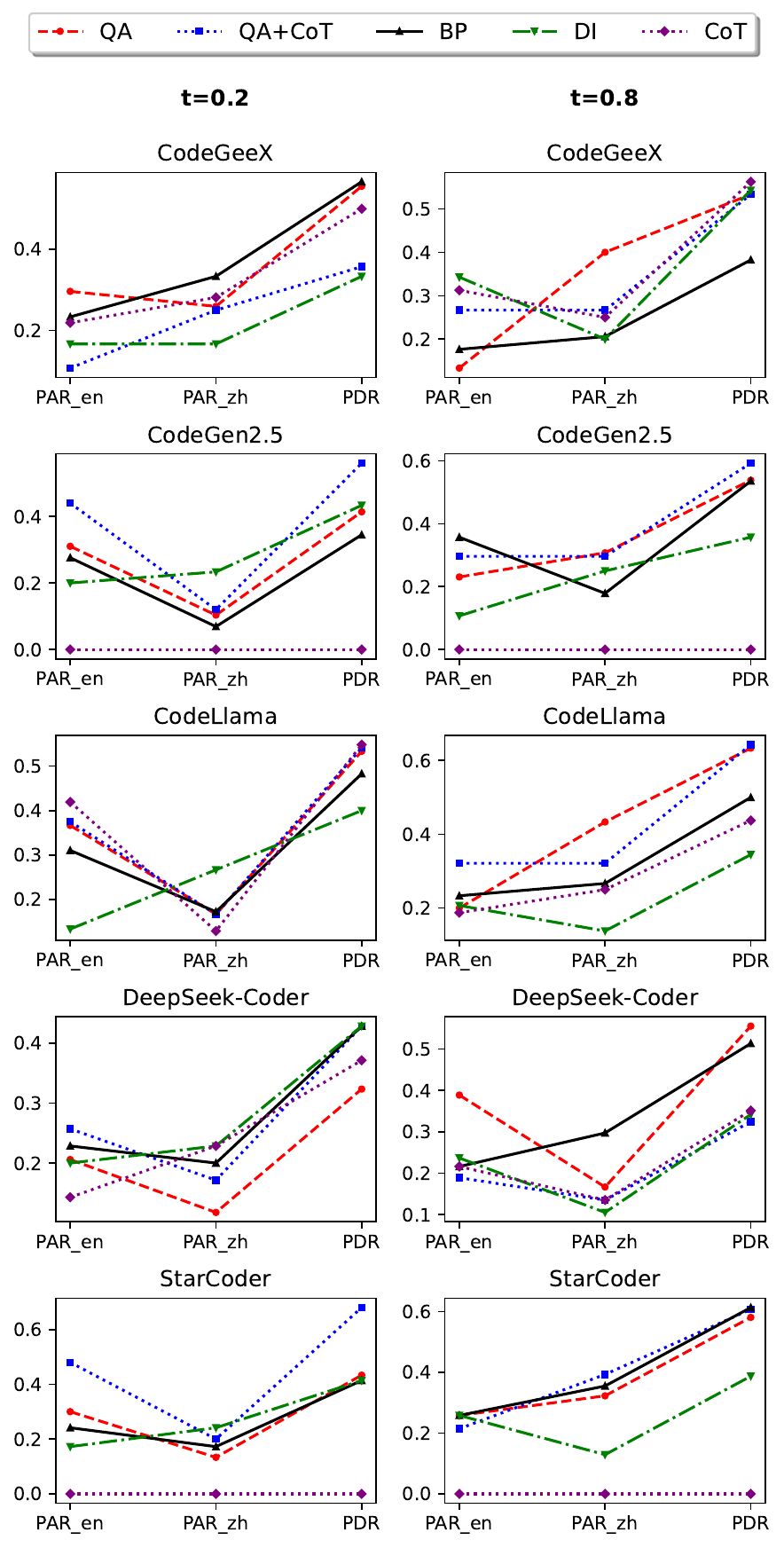}
    }
    \caption{Impact of Prompting on Bilingual Code Generation.}
    \label{fig:impact_prompting}
  \end{figure*}

\subsection{RQ3: The Impact of Prompting}\label{sec:rq3}


\begin{figure*}[!h]
  \centering
  \includegraphics[width=0.75\linewidth]{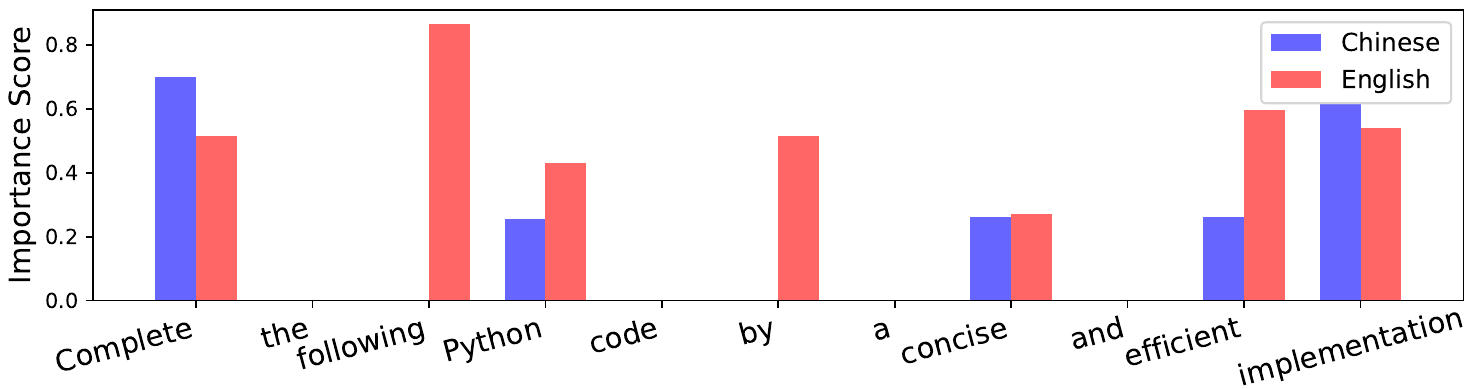}
  \caption{A visualization of the importance distribution on input prompts of DeepSeeker. \footnotesize{The importance score of each word ranges from 0 to 1, where the missing bars indicate a value of 0.}}  \label{fig:xai}
\end{figure*}


\noindent\textbf{Design.}
In this RQ, we investigate the impact of different prompting methods on the quality of LLM-generate code in bilingual programming questions. 
As shown in ~\autoref{fig:prompting}, in previous experiments, we directly fed those questions into LLMs. 
Now, we introduce different prompting paradigms to provide clearer and more specific instructions, aiming to improve the quality of code generation.
First, we experiment with a basic prompt, which is commonly utilized by some instruction-tuned LLMs to enhance generation capability~\cite{li2023starcoder}. 
Building upon this, we further attempt to introduce quality-aware information into the prompt, explicitly instructing the model to generate ``concise and efficient'' code, to improve the correctness and efficiency of the generated code.
Moreover, the Chain-of-Thought (CoT)~\cite{wei2022chain} prompting method has recently been widely proven to significantly improve the quality of generation results. 
A typical example is adding statements like ``Let's think step by step'' in the tail of the prompt, guiding the model to perform step-by-step thinking and reasoning. 
Therefore, we also incorporate the CoT prompting method in our experiments.
Finally, to comprehensively consider both code quality and the logical coherence of the generation process, we experiment with the combination of quality-aware information and the CoT prompting method.

\noindent\textbf{Results.}
With our observation, we find that \textit{
different prompting methods have varying impacts on the correctness and efficiency of LLM-generated code in bilingual programming questions}.
~\autoref{fig:impact_prompting} presents the impact of different prompting methods on the correctness and efficiency of LLM-generated code based on our proposed metrics.
From a correctness perspective, using these diverse prompts may lead to a decrease in the ability of LLMs to generate correct code in both Chinese and English languages compared to directly inputting programming questions, especially for CodeGeeX and StarCoder. 
From an efficiency standpoint, prompts also have varying degrees of influence. 
For example, at $t=0.2$, it can be observed that diverse prompts significantly bias code generation towards English in models like CodeGen2.5, CodeLlama, DeepSeek-Coder, and StarCoder. 
Moreover, the impact of increasing temperature \(t\) on the correctness and efficiency of code generation varies across different prompts on LLMs.
For example, as \(t\) changes, the code generation correctness of prompts ``QA'' and ``QA+CoT'' fluctuates significantly, while prompts 'DI' and 'BP' show minimal sensitivity to changes in \(t\).
Overall, different prompting strategies affect the quality differences in code generation by LLMs in bilingual contexts, and this difference is often influenced by temperature, with varying effects on different LLMs.

To simpler and more straightforward, we compared four prompting methods against direct input (DI) across all conditions. At temperature 0.2, only BP and QA consistently improved $PAR_{en}$ across models. At temperature 0.8, only QA+CoT and QA showed consistent improvements in $PAR_{zh}$ instead of $PAR_{en}$. These results demonstrate that \textit{no prompting method maintained consistent improvements across all experimental settings}.

Furthermore, we attempt to explore why LLMs exhibit different behaviors to the same prompting method instruction when the tasks are described in English and Chinese. 
Taking the Quality-aware Prompt(QA) as an example, we are curious whether LLMs pay significantly different attention to the key information terms ``concise'' and ``efficient'' under the two language inputs.
More specifically, we calculate the importance scores of each word in the input prompt to LLMs through the perturbation-based interpretability technique, which are considered to produce the best attention alignment to human programmers~\cite{kou2024large}. 
We make an observation with DeepSeek-Coder on specific string tasks, as shown in~\autoref{fig:xai}. Our findings reveal that when using Chinese task descriptions, the importance score of ``concise'' slightly decreases, while the score of ``efficient'' significantly drops.

\section{Conclusions}
\label{sec:conclu}

In this paper, we conduct a pioneering empirical study on the linguistic bias towards LLMs for code generation. With English and Chinese as the focused languages, we design a differential evaluation framework to assess code quality ranging from correctness to efficiency. Our dataset comprises 52 bilingual parallel programming tasks, and we evaluate the code generated by eight models from five popular open-source LLM families as well as GPT-3.5-Turbo and GPT-4. Our findings underscore significant quality disparities between the code generated by LLMs for English and Chinese programming questions, thereby revealing the existence of linguistic bias. This work calls for greater attention from the software engineering community to ensure that advanced productivity tools provide equal opportunities to developers worldwide.

\section*{Limitations and Ethics}
\label{sec:limitations}

\noindent\textbf{Limitations.} Our study faces three main limitations. 
First, our dataset is constrained to 52 English-Chinese programming question pairs, which may not fully represent the diversity of real-world programming tasks and language varieties. 
Second, while we evaluated 10 popular models from 6 different sources (5 open-source families and one commercial provider), this selection cannot encompass all rapidly evolving new LLMs. Third, our metrics focus primarily on code correctness and efficiency, excluding other important aspects such as readability and maintainability. To address these limitations, we have made our dataset and evaluation framework publicly available to facilitate future research incorporating additional programming tasks, languages, models, and quality metrics.
Notably, we believe our evaluation have provided compelling evidence of linguistic bias risks in LLM-based code generation.

\noindent\textbf{Ethics.} In this study, we aim to advance social equity by exposing disparities in code generation quality across different natural languages. By highlighting these linguistic biases in LLMs, we seek to raise awareness within the research community and advocate for their mitigation, ultimately contributing to a more inclusive technological landscape.




\bibliography{ref}

\appendix

\section{Appendix}
\label{sec:appendix}

\subsection{Example of Our Test Cases}
\label{sec:appendix_examples}

\begin{figure*}[!t]
    \centering
    \includegraphics[width=\linewidth]{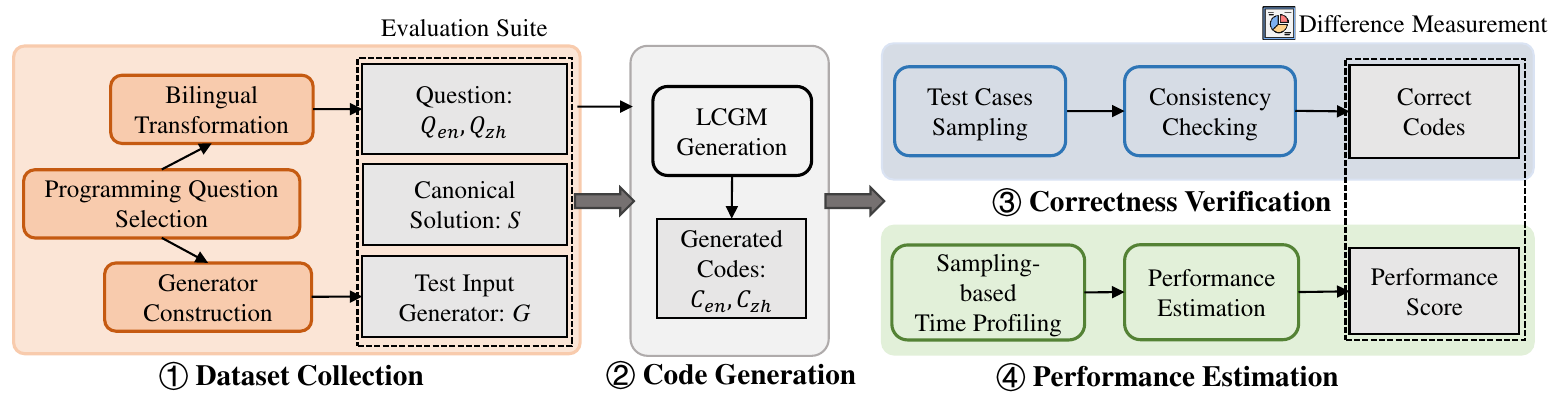}
    \caption{The workflow of our evaluation. }

    \label{fig:overview}
\end{figure*}

\begin{figure*}[tbp]
    \centering
    \includegraphics[width=0.9\linewidth]{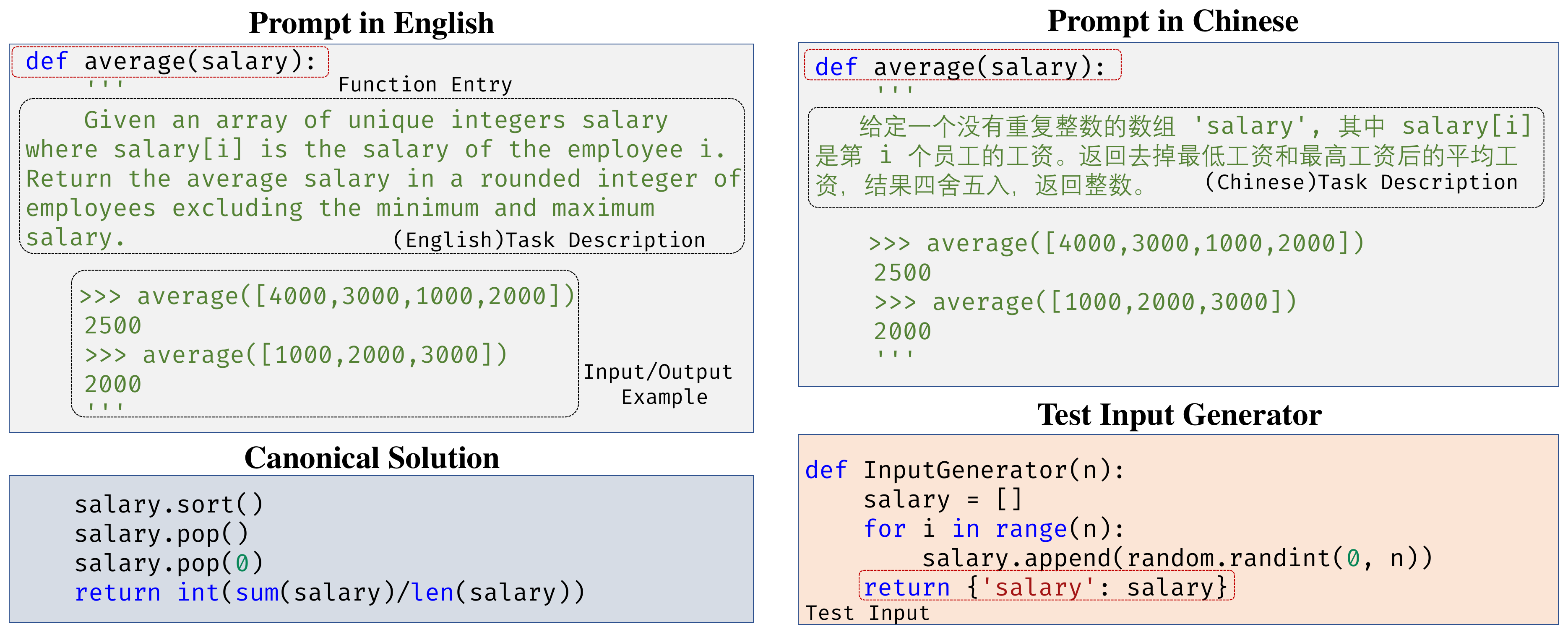}
    \caption{An example in our test suite.}

    \label{fig:examples}
\end{figure*}

As shown in~\autoref{fig:overview}, our test suite can be formally represented as $\mathcal{D} = \langle Q, S, G \rangle$, where $Q$ is the programming question, $S$ is the canonical solution, and $G$ is the test input generator. 
$Q$ consists of a task description, an entry point function name, and reference input/output. 
For each question \(q \in Q\), we have two bilingual parallel versions, \(q_{en}\) and \(q_{zh}\), presented in English and Chinese, respectively.
$S$ is the canonical solution of the given programming questions, which is used to produce ground truth outputs for correctness verification.
$G$ is a crucial component, serving as the input generator. 
\autoref{fig:examples} shows an example of a programming question in our test suite.
It takes an input size parameter $n$ that determines the dimensionality of the generated variables. These variables are tailored to act as unit test inputs for the specific programming questions.
By varying the size parameter, $G$ can generate a wide range of diverse inputs, from small and simple cases to large and complicated ones. 

\subsection{Adaptive Maximum Input Size Selection}
\label{sec:appendix_time_series}
In \autoref{sec:design_performance}, we mention that we propose an adaptive mechanism to seek and select the appropriate maximum input size to ensure that the captured execution time is within a reasonable range that is not too small to cause too much randomness impact, while not too large to bring unacceptable computational consumption. 
We will introduce the detailed process.
Initially, when constructing the input generator, we empirically set an initial maximum input size $N_{max}$ for each programming question, taking into account the complexity of the canonical solution.
This value is set to be sufficiently large, allowing us to divide it into many segments to sample different input sizes.
However, a problem arises as the LLM-generate code could have a higher complexity than the canonical solution, leading to excessively long execution times and making the overall evaluation time prohibitively expensive. 
To address this problem, we design an adaptive mechanism for adjusting the maximum input size.
Following HumanEval, we first set a maximum time limit $T_{max}$ for code execution.
Subsequently, we examine the execution of each LLM-generated code to determine if it triggers a timeout when processing inputs of size $N_{max}$. In the event of a timeout, we iteratively decrease $N_{max}$ by half until the execution satisfies the $T_{max}$ constraint.
With the maximum input size $N_{max}$, we divide it into $seg$ equal parts and obtain an increasing sequence of input sizes:
\[N = {n_i | n_i = i \cdot \frac{N_{max}}{seg}, i = 1, 2, \dots, seg}\]
For each input size $n_i$, we perform $k$ samplings, record the execution time of each run, and take the average as the execution time estimate $t_i$ under that input size. This way, we obtain a set of ``input size-execution time'' pairs:
\[NT = {(n_i, t_i) | n_i \in N, t_i = \frac{1}{k}\sum_{j=1}^{k}t_{ij}}\]
where $t_{ij}$ represents the execution time of the $j$-th sampling when the input size is $n_i$.

\subsection{Detailed Metrics}
\label{sec:appendix_metrics}
\subsubsection{Correctness}
The correctness is measured through the correctness verification procedure. 
The function \(CorVer\) takes a problem $q$ as input and returns a boolean value. It returns $True$ if the problem $q$ has been correctly solved by the target LLM, and $False$ otherwise.
Thus, we define the following evaluation metrics:

\noindent\(\bullet\) $\mathbf{CR_{en}}$ (Correctness Rate in English): This metric represents the proportion of programming questions in English for which the target LLM can generate correct code. The specific calculation of \(\text{CR}_{en}\) can be formalized as:
\[\frac{|{q \in Q: \text{CorVer}(q_{en}) = True}|}{|Q|}\]

\noindent\(\bullet\) $\mathbf{CR_{zh}}$ (Correctness Rate in Chinese): Similarly, this metric represents the proportion of programming questions in Chinese for which the target LLM can generate correct code. The \(\text{CR}_{zh}\) is calculated as:
\[
\frac{|{q \in Q: \text{CorVer}(q_{zh}) = True}|}{|Q|}
\]

\noindent\(\bullet\) $\mathbf{CR_{bi}}$ (Correctness Rate in Bilingual): This metric represents the proportion of programming questions for which the target LLM can generate correct code in both languages. The \(\text{CR}_{bi}\) is calculated as:
\[
\frac{\left|\{q \in Q : \text{CorVer}(q_{en}) \wedge \text{CorVer}(q_{zh}) = \text{True}\}\right|}{|Q|}
\]


\noindent\(\bullet\) $\mathbf{CDR}$ (Correctness Difference Rate): This metric represents the proportion of programming questions for which the target LLM can generate correct code only for one specific language. The \(\text{CDR}\) is calculated as:
\[\frac{|{q \in Q: \text{CorVer}(q_{en}) \oplus \text{CorVer}(q_{zh}) = \text{Ture}}|}{|Q|}\]
where \(\oplus\) is the XOR operator.

\subsubsection{Efficiency}
As aforementioned, we evaluate the efficiency differences on the programming questions subset where LLMs can generate correct solutions for both bilingual inputs.
Firstly, we define the subset of correctly solved programming questions as \(Q^\prime \subseteq Q\), which satisfies
$$\forall q \in Q^\prime, \text{CorVer}(q_{en}) \wedge \text{CorVer}(q_{zh}) = \text{True}$$
Besides, we introduce  $SComplex$ as a function that inputs a programming question $q$ and returns the score of the estimated complexity for the LLM-generated code. 
The score is a value from 1 to 7, with lower scores indicating higher efficiency in terms of actual code execution.
The measurement metrics for efficiency are as follows:

\noindent\(\bullet\) $\mathbf{PAR_{en}}$ (Performance Advantage Rate in English): This metric represents the proportion of programming questions in $Q^\prime$ for which the LLMs generate code with better estimated time complexity (lower $SComplex$ value) in English compared to Chinese. The \(\text{PAR}_{en}\) is calculated as:

\[\frac{|{q \in Q^\prime: \text{SComplex}(q_{en}) < \text{SComplex}(q_{zh})}|}{|Q^\prime|}\]

\noindent\(\bullet\) $\mathbf{PAR_{zh}}$ (Performance Advantage Rate in Chinese): This metric represents the proportion of programming questions in $Q^\prime$ for which the LLMs generate code with better estimated time complexity (lower $SComplex$ score) in Chinese compared to English. The \(\text{PAR}_{zh}\) is calculated as:
\[\frac{|{q \in Q^\prime: \text{SComplex}(q_{zh}) < \text{SComplex}(q_{en})}|}{|Q^\prime|}\]

\noindent\(\bullet\) $\mathbf{PDR}$ (Performance Difference Rate): This metric represents the proportion of programming questions in $Q^\prime$ for which the LLMs generate code with significantly different estimated time complexities (different $SComplex$ scores) between English and Chinese. The \(\text{PDR}\) is calculated as:
\[\frac{|{q \in Q^\prime: \text{SComplex}(q_{en}) \neq \text{SComplex}(q_{zh})}|}{|Q^\prime|}\]

\subsection{Target LLMs Details}
\label{sec:appendix_models}
The LLM landscape is rapidly evolving, with a constant influx of new models, including both open-source frameworks and commercial APIs. Given the pace of these developments, it is impractical to conduct an exhaustive study of all available models. Therefore, our goal is to select a representative subset of models to demonstrate the presence of linguistic bias.
In our study, we choose the following five kind of open-source LLMs based on three criteria: (1) their support for both English and Chinese programming tasks, as demonstrated by strong performance on existing benchmarks; (2) the inclusion of diverse and representative model families or architectures to ensure the generalizability of findings related to linguistic bias; and (3) due to GPU resource limitations, requiring that models do not exceed 34 billion parameters.
\begin{itemize}
    \item CodeGen2.5~\cite{Nijkamp2023codegen2}: An extension of the popular CodeGen2 family, CodeGen2.5, with 7B parameters, is claimed by its developers to perform on par with 15B parameter LLMs, following additional training on Python. It was developed by Salesforce.

    \item StarCoder~\cite{li2023starcoder}: Trained on 80+ programming languages, StarCoder supports a novel combination of capabilities and architectural features unavailable in other open LLMs. We choose the 13B version.
    
    \item CodeGeeX~\cite{zheng2023codegeex}: CodeGeeX is a multilingual code generation model implemented based on the ChatGLM2 architecture. It has 6.7B parameters and particularly supports bilingual input in English and Chinese.
    
    \item CodeLlama~\cite{roziere2023CodeLlama}: Developed by Meta, CodeLlama is a foundational model with various parameter size versions. We consider the 7B, 13B, and 34B versions. Unless otherwise specified, any subsequent references will refer to the 7B version by default.
    
    \item DeepSeek-Coder~\cite{guo2024deepseek}: This model comprises a series of code language models trained from scratch on both 87\% code and 13\% natural language in English and Chinese. It supports project-level code completion and infilling. We consider the 7B and 33B versions.
\end{itemize}
Meanwhile, our evaluation also covers OpenAI's GPT-3.5-Turbo and GPT-4, two state-of-the-art commercial models~\cite{gpt3.5-oepnai}.
Thus, our evaluation involves a total of ten models. However, due to computation resources and costs, we only consider the five small versions of the open-source models when conducting some of the additional investigations.

\subsection{Parameter Configuration}
\label{sec:appendix_parameters}

In our experiments, the maximum time limit $T_{max}$ is 6 seconds, the number of sampling $k$ is 100, and the number of segments $seg$ is 50.
Another important parameter when utilizing LLMs is the temperature ($t$), which typically ranges from 0 to 1. 
Lower temperature values result in more deterministic and concentrated outputs, while higher values lead to more diverse and exploratory generations. 
We select two common temperature values, 0.2 and 0.8, to better observe the impact of this parameter on the quality of the generated code~\cite{OpenAI-API-doc}.

Our evaluation framework incorporates several configurable parameters. The calibration of these parameters is crucial, as excessively low values may introduce undue randomness, while overly high values can significantly increase computational time. To address this trade-off, we have meticulously adjusted these parameters.
To alleviate concerns regarding the reliability of our evaluation with respect to parameter selection, we conducted convergence experiments. These experiments aimed to verify the stability of our results across different parameter configurations. 
We randomly select 30 code snippets generated for different programming questions. We try maximum time limits \(T_{max}=6,8,10\), the number of samples \(k=100,150,200\), and the number of segments \(seg=50,100,200\). 
That means we increase the parameters used in the paper to observe the consistency of complexity estimation results. The results show that all 30 outcomes are consistent as Tmax changes, and 29 outcomes are consistent with changes in \(k\) and \(seg\). The only inconsistency arises from slight errors in quadratic fitting. 
Overall, our chosen parameter settings exhibit strong stability in the assessment of results. The significant consistency observed across different parameter configurations enhances the reliability of our evaluation framework and supports the validity of our conclusions.



\end{document}